\documentclass[12pt,notoc]{JHEP3}
\usepackage{amsmath,amssymb,euscript,array,mathrsfs}
\newcommand{\startappendix}{
\setcounter{section}{0}
\renewcommand{\thesection}{\Alph{section}}}
\newcommand{\Appendix}[1]{
\refstepcounter{section}
\begin{flushleft}
{\large\bf Appendix \thesection: #1}
\end{flushleft}}
\usepackage{epsfig}

\def\N{{\cal N}}
\def\tr{{\rm tr}}

\def\Tr{{\rm Tr}}

\newcommand{\Dslash}{D\mkern-11.5mu/\,} 
\newcommand{\delslash}{{\partial\mkern-9mu/}}

\newcommand{\nfour}{\ensuremath{\mathcal{N}\!=\!4\,}}

\def\Dbarslash{\,\,{\raise.15ex\hbox{/}\mkern-12mu {\bar D}}}
\def\Dslash{\,\,{\raise.15ex\hbox{/}\mkern-12mu D}}
\def\delslash{\,\,{\raise.15ex\hbox{/}\mkern-9mu \partial}}
\def\delbarslash{\,\,{\raise.15ex\hbox{/}\mkern-9mu {\bar\partial}}}

\newcommand{\EQ}[1]{\begin{equation} #1 \end{equation}}
\newcommand{\AL}[1]{\begin{subequations}\begin{align} #1
\end{align}\end{subequations}}
\newcommand{\SP}[1]{\begin{equation}\begin{split} #1
\end{split}\end{equation}}

\title{${\mathbb Z}_N$ Domain walls in hot $\N=4$ SYM at weak and strong coupling}  
\author{Adi Armoni, S. Prem Kumar and Jefferson M. Ridgway\\
{\it 
Department of Physics, Swansea University, \\ Singleton
  Park,Swansea\\SA2 8PP, U.K. 
}\\ 
E-mail: \email{a.armoni}, \email{s.p.kumar}, \email{pyjr@swansea.ac.uk}
} 
\abstract{We study the tensions of domain walls in the deconfined
  phase of $\N=4$ SUSY Yang-Mills theory on ${\mathbb R}^3\times S^1$, 
  at weak and strong coupling. We calculate
  the $k$-wall tension at one-loop order and find that it is
  proportional to $k(N-k)$ (Casimir scaling). The two-loops analysis
  suggests that Casimir scaling persists to this order. The strong
  coupling calculation is performed by using the AdS/CFT
  correspondence. We argue that the $k$-wall should be identified with
  an NS5-brane wrapping an $S^4$ inside $S^5$ in the
  AdS-Schwarzschild$\times S^5$ background in Type IIB string theory. 
  The tension at strong
  coupling is compared with the weak coupling result. We also compare
  our results with those from lattice simulations in pure
  Yang-Mills theory.}

\begin{document}
\section{Introduction}
\label{sec:intro}

In $SU(N)$ gauge theories with adjoint matter the Polyakov-loop
expectation value serves as an order parameter for
the deconfinement transition \cite{Yaffe:1982qf, Svetitsky:1982gs}. 
In the confining phase, the centre of the
gauge group, ${\mathbb Z}_N$ for $SU(N)$, is unbroken and the Polyakov loop
expectation value is zero. In the deconfined phase on the other hand, 
the centre is spontaneously broken and the Polyakov loop expectation value
is an element of the centre, $P= \exp(i 2\pi k/N)$. 
When the discrete ${\mathbb Z}_N$ symmetry is spontaneously broken there
exist domain walls that interpolate between the various vacua. The
``fundamental'' domain wall interpolates between two adjacent
vacua. More generally, a domain wall which interpolates between one
vacuum $P= \exp (i 2\pi l/N)$, and another with 
$P=\exp i 2 \pi (l+k)/N$, is called
a $k$-wall. By definition, it is obvious that all the properties of
the $k$-wall (e.g. tension or width) should be invariant under $k
\rightarrow k+N$. Charge conjugation implies invariance under $k
\rightarrow N-k$ as well. 

This paper is devoted to the study of domain walls in the deconfined phase of 
${\cal N}=4$ SUSY Yang-Mills theory on ${\mathbb R}^3\times S^1$,
with supersymmetry breaking boundary conditions for fermions. 
${\cal N}=4$ SYM is a useful toy 
model, since the domain walls can be analyzed at both weak and strong
coupling. The weak coupling calculation is performed by a perturbative
analysis, while the strong coupling analysis is performed by using the
AdS/CFT correspondence. 

Not much is known about deconfined phase domain walls of the $\N=4$ theory. 
The literature that
we will briefly review momentarily, is almost strictly devoted to
domain walls in pure Yang-Mills theory. The tension of the domain
walls in pure Yang-Mills theory was computed at high temperature and
consequently, weak coupling, at one-loop order, in the seminal papers of
Bhattacharya {\it et al.} \cite{Bhattacharya:1990hk,Bhattacharya:1992qb}. The
result is 
\EQ{
{\cal T} _ k = k(N-k) {4\pi ^2 \over 3 \sqrt 3} {T^2 \over \sqrt
  {g_{\rm YM}^2(T) N}} \label{YM-one-loop-result} \, . 
}
where $g_{\rm YM}(T)$ is the running gauge coupling at temperature $T$.
The above result \eqref{YM-one-loop-result} was extend to two-loop
order in \cite{Giovannangeli:2001bh}. In particular, the tension up to
two-loop order still exhibits a Casimir scaling, namely ${\cal T} _ k
\propto k(N-k)$. The three-loop analysis \cite{Giovannangeli:2002uv}
reveals a deviation from Casimir scaling. The tension of domain walls
was also measured by lattice simulations
\cite{deForcrand:2004jt,Bursa:2005yv,deForcrand:2005rg}. The
conclusion of \cite{deForcrand:2004jt,Bursa:2005yv,deForcrand:2005rg} is
that Casimir scaling holds within the measurement error to
low-temperatures (close to $T_c$), where perturbation theory is not valid
anymore. Other aspects of deconfining phase domain walls and of particular
significance for us,
the relation between domain walls and spatial 't Hooft loops were
investigated in 
\cite{KorthalsAltes:1999xb}
\footnote{The physical status of
  these domain walls for the real time hot gauge theory in 3+1
  dimensions has been questioned in \cite{smilga} and other works. We
  will, however, only view the ${\mathbb Z}_N$ walls as interfaces in
  the effective 3 dimensional theory on ${\mathbb R}^3\times S^1$
  where the circle is interpreted as a spatial direction.}. 

Within the AdS/CFT framework, domain walls in the deconfined phase 
of ${\cal N}=4$ SYM, in the large-$N$ limit and strong 't Hooft coupling,
were discussed by Aharony and Witten \cite{Aharony:1998qu}. They
proposed that the fundamental domain wall should be  
identified with a D1-brane, reinforcing the connection between 
${\mathbb Z}_N$ walls and spatial 't Hooft
loops. Their work is reviewed in Section
\ref{strong-coupling}.  

The purpose of this paper is two-fold: (i) To carry out the
perturbative analysis of
\cite{Bhattacharya:1990hk,Giovannangeli:2001bh} for the ${\cal
  N}=4$ SUSY Yang-Mills theory. (ii) To extended the analysis of
\cite{Aharony:1998qu} to the case of a $k$-wall. To this end, we
propose that the $k$-wall is an NS5-brane (or a D5-brane in the S-dual
theory) that wraps an $S^4$ inside the $S^5$ in the Type IIB string
theory geometry, dual to the strongly coupled $\N =4$ theory. The
geometry dual to the Euclidean field theory on ${\mathbb R}^3\times
S^1$ is the Euclidean AdS-Schwarzschild black hole.

Our main results are as follows: at the one-loop order we find that the
tension of the $k$-wall in the $\N=4$ theory is 
\EQ
{
{\cal T} _ k = k(N-k)\frac {4\pi^2T^2}{\sqrt{3\lambda} } \frac{\sqrt 2}{5} (3\sqrt 3 -2) \,;
  \qquad\quad \lambda={\small g^2_{\rm YM}N}\ll 1
\label{N4-one-loop-result} 
}

We extend our perturbative analysis beyond one-loop and argue that up
to two loops the tension exhibits a Casimir scaling as in the pure YM
case. Although we did not carry out the three-loop calculation, we do
not expect that the Casimir scaling behaviour will hold beyond two
loops. 

In the limit of strong 't Hooft coupling $\lambda\gg1$, in the large
$N$ limit, we find that the IIB supergravity dual yields a domain
wall tension which is no longer proportional to $k(N-k)$. Instead we find
that 
\EQ{
{\cal T} _k =\tfrac{4}{3} \pi \;N^2  \;{T^2 \over \sqrt
  {\lambda}}\; \sin ^3 \alpha _k \, , 
\qquad\quad\tfrac{1}{2}\sin 2 \alpha _k- \alpha _k = 
\pi (\tfrac{k}{N} -1). \label{sugra-result} 
}
 This formula is valid when $N \rightarrow \infty$ while $k/N$ is kept
 fixed. It is worth mentioning that \eqref{sugra-result} is extremely 
well approximated, within a maximum error of $2\%$, by a different
expression 
 \cite{Armoni:2006ux} 
\EQ{
{\cal T} _k \approx 2 \pi\, N^2  {T^2 \over \sqrt {\lambda} }\left
  ( \sin \pi \tfrac {k}{N} - \tfrac{1}{3} (\sin  \pi \tfrac{k}{N}
  )^{\frac{3}{2}} \right ) \, .  
} 

The paper is organized as follows: Section 2 is devoted to preliminary
definitions and review. 
In Section 3 we carry out a detailed one-loop calculation
of the $k$-wall tension. In Section 4 we argue that the Casimir
scaling behavior should persist at two-loop. In Section 5 we calculate
the wall tension at strong 't Hooft coupling by using supergravity. In
section 6 we discuss out results.

\section{Preliminaries}

\paragraph{}
The order parameter for the deconfinement transition in $SU(N)$ gauge
theories, with matter transforming in the adjoint representation,
is the Polyakov
loop. It is strictly only an order parameter in the Euclidean thermal
theory and is defined as the Wilson line around the (compactified) 
Euclidean time
direction, 
\begin{equation}
P(x) = \frac 1 N {\Tr}\, {\cal P}\,
\exp \left( i \,g_{\rm YM} \int^\beta_0 A_0
  (x,t) dt \right),\, \beta ={1\over T}.
\label{eq:wilson_loop_1}
\end{equation}
At high temperatures, across the deconfinement transition the Polyakov
loop acquires an expectation value, breaking the ${\mathbb Z}_N$
center symmetry and signalling a deconfined phase 
\cite{Yaffe:1982qf, Svetitsky:1982gs}. The spontaneous breaking of the
${\mathbb Z}_N$ symmetry leads to $N$ distinct phases labelled by the
$N$ roots of unity. When the Euclidean gauge theory on ${\mathbb
  R}^3\times S^1$ is viewed as a three dimensional effective field theory, one
can imagine domain walls interpolating between regions with different
expectation values of $P$. As first established in the works of
\cite{Bhattacharya:1990hk, Bhattacharya:1992qb, korthalsaltes}, 
the explicit profiles
of the ${\mathbb Z}_N$ interfaces and their
tensions can be computed at weak coupling, utilizing high temperature
perturbative and semiclassical methods. The ${\mathbb Z}_N$ domain
wall is like an instanton in a one dimensional effective theory
describing the profile perpendicular to the interface. The
perturbative description works because the thickness of the domain
wall is set by $(g_{\rm Y M} N^{\frac{1}{2}} T)^{-1}$, the Debye
screening scale.  

In what follows, we will review ({\it cf.} \cite{korthalsaltes}) 
the semiclassical description 
of the ${\mathbb Z}_N$ interfaces and apply it to the $\N=4$ theory at high
temperature and weak gauge coupling. The first step in this
computation involves parametrizing the varying expectation value of the
Polyakov loop across the domain wall,
by a classical background value for $A_0$.
Thus the temporal gauge field $A_0$ can be split into
classical and quantum parts, 
\EQ
{
A_0= A_0^{\rm cl} +A_0^{\rm qu}.
}
\subsection{The fundamental wall or $k=1$}
In the three dimensional effective theory, the domain interface
 can be thought of as a string-like object moving through
 time. Taking this interface to span the $x-y$ plane, 
 the different vacua then sit at different values of $z$. 
Focusing on the classical, or background field part, we pick an ansatz where 
$A_0^{\rm cl}$ is expressed in terms of the diagonal traceless
generator $t_N$,  
\AL
{
&A_0^{\rm cl} = \frac{2 \pi \,T}{g_{\rm YM} N} \,{t_N \over
  B(N)}\, q(z)\,;\quad
t_N= B(N)\,\, \mathrm{Diag}[\underbrace{1,1,1,\dots,1}_{N-1\,
  \mathrm{entries}},1-N]\label{k1}\\ 
&B(N)=\frac{1}{\sqrt{2\,N\,(N -1)}}.
\label{norm}
}
The spatial components of the gauge field are chosen to 
vanish, $A_i=0$ for $i=1,2,3$. Importantly, the profile function 
$q$ parametrizes the $N$
different vacua. This can be seen more explicitly by evaluating the
Polyakov loop order parameter with the above ansatz on a  constant 
background field $A_0^{\rm cl}$
(the explicit $z$ 
dependence will be dealt with later),
\begin{equation}
P = \frac 1N \left[(N-1)e^{\frac{2\pi i q}{N}} + e^{\frac{2\pi i q
      (1-N)}{N}}\right].
\end{equation}  
Allowing $q$ to take on one of $N$ integer values,
$q=0,1,2,\dots,N-1$, we can scan through each of the $N$ vacua labelled by 
$P=e^{\frac{2\pi i q}{N}}$. 
As
 the vacua are labelled by integer values of $q$, the domain
 interface itself will be characterized by non-integer values of the profile 
function $q(z)$ interpolating between two vacua.

The ansatz \eqref{k1} can be used to describe a wall that interpolates
between two ${\mathbb Z}_N$ phases labelled by consecutive integers.
 Since all the vacua are
 physically equivalent, to discuss the wall tension, we may focus on 
the interface
 between $q=0$ and $q=1$ without loss of generality. Therefore, we may
 think of $q(z)$ as interpolating between $q(z=0)=0$ and $q(z=L)=1$,
 where $L$ is the extent or thickness of the interface between
 neighbouring vacua.   

Up to this point, the walls under scrutiny have been the 
fundamental or $k=1$ walls. To consider walls between vacua with
multiple ${\mathbb Z}_N$ ``charge''
 difference, $k$, the ansatz for $A_0^{\rm cl}$ must be
 modified as below.
\subsection{The $k$-wall ansatz}
We first set out our conventions for the $N^2-1$ generators of 
$SU(N)$. Separating them into the $N-1$ diagonal generators of the
Cartan subalgebra and $N(N-1)$
off-diagonal or ladder generators, the Cartan elements are
of the form, 
\begin{equation}
t_{\rm diag} \equiv t_i =B(i)\,\,
\mathrm{Diag}[\underbrace{1,1,1,\dots,1}_{i-1\, \mathrm{entries}},1-i,
\underbrace{0,0,\dots,0}_{N-i\, \mathrm{entries}}] \,\qquad i\in[2,N]
\label{eq:gen_diag_general}
\end{equation}
The normalization $B(i)$ as defined in \eqref{norm} ensures that
\begin{equation}
\Tr(t_i \,t_j )= \frac 12 \delta_{ij}.
\end{equation}
For every Cartan generator $t_i$, we can define 
$2(i-1)$ ladder generators, $t_{ij}$ with one non-zero element:
\begin{equation}
t_{ij}^{mn}=\frac 1{\sqrt{2}} \delta_i^n \delta_j^m\,,\qquad
j\in[1,i-1].\label{eq:gen_ladder}
\end{equation}
$t_{ij}$ provides the off-diagonal
generators with the non-zero matrix elements in the upper right half, 
while the lower left off-diagonal generators are given by the transpose,
$t_{ji}$. The off-diagonal ladder generators are normalized in the
following way 
\begin{equation}
\Tr(t_{ij}\,t_{j^\prime i^\prime}) = \frac 12\, \delta_{ii^\prime}\,
\delta_{jj^\prime}.
\end{equation}
The algebra of the generators simplifies significantly in
this basis, with the only non-vanishing  commutators being 
\begin{equation}
[ t_i, t_{ij}] = N B(i)\, t_{ij}\,; \,\,\,\,\,\, [ t_i, t_{ji}] = - N
B(i)\, t_{ji}.
\label{eq:gen_ladder_commutators}
\end{equation}

Returning to the idea of an interface between two vacua labelled by
generic integers, a so called  $k$-wall, a modified ansatz is
required.  As a $k$-wall is an
interface between two vacua with a ${\mathbb Z}_N$
charge difference $k$, the ansatz 
for $A_0^{\rm cl}$  is chosen to be proportional to the hypercharge
matrix $Y_k$
\begin{equation}
A_0^{\rm cl}  = \frac{2 \pi T}{g_{\rm YM} N } \,q(z)\, Y_k.
\end{equation}
where $Y_k$ is defined as
\begin{equation}
Y_k\equiv \,\mathrm{Diag}[\,\underbrace{k,k,k,\dots,k}_{N-k\,
  \mathrm{entries}},\underbrace{k-N,k-N,\dots,k-N}_{k\,
  \mathrm{entries}}\,]\,,\qquad k\in[1,N]
\label{hypercharge}
\end{equation}

As with $t_N$, $Y_k$ is
traceless, and the resulting ansatz
will be symmetric under $k\leftrightarrow N-k$ which is required by
${\mathbb Z}_N$ invariance of the theory.

Applying these modifications to the order parameter, the
role of $q$ is now clear. Previously, for $k=1$, the parameter $q$
defined each vacuum individually when integer valued, and 
non-integer values of $q$ characterized a point within an interface. For the
$k$-walls with $k>1$,
it is no longer $q$, but the product $k q$ that specifies a given vacuum for
integer values of $q$; $q$ now becomes a parameter varying across the $k$-wall,
as before from $q(0)=0$ to $q(L)=1$. The Polyakov loop 
order parameter, for this ansatz is
\begin{equation}
P=\frac 1N \left[(N-k)e^{\frac{2\pi i }{N}k\, q}+k \,e^{\frac{2\pi i
    }{N}(k-N)\, q}\right] 
\end{equation}
with $P=1$ at $q=0$ as before, and  
$P=e^{\frac{2\pi i }{N}k}$ when $q=1$.
\section{$k$-Wall Tension in \nfour \,SYM}
\label{sec:n=4_kwall}
Having specified the ansatz for the $k$-wall solution, our 
aim now is to determine the tension of the
$k$-wall in $\N=4$ SYM. The basic idea is to insert the classical or
background profile for the ${\mathbb Z}_N$ instanton and use weak
coupling to expand in the quantum fluctuations around the background
configuration. The quantum fluctuations induce a one-loop effective
potential for $q(z)$, which is then used to determine the wall
solution and its tension. Crucially, it is necessary to ensure
self-consistently that the resulting configuration can be understood
at weak coupling. We will elaborate on this subsequently.
 
At the classical level, as shall be shown,
there is no interface solution, so one-loop effects must be
included. The only terms in the action of $\N=4$ SYM that we need to
focus attention on, are those that involve the interactions of the background
gauge configuration $A_0^{\rm cl}$ with quantum fluctuations:
\begin{equation}
S = \int\!\! d^3\!x \,\int_0^\beta\!\!d\tau \,\!\left\{\Tr\left[\frac
    12 (F_{\mu\nu})^2\right] +  \Tr\left[\sum_{A=1}^4
    \,\overline{\psi}_A D\!\!\!\!\slash\,\,\,\psi_A\right] +
  \Tr\left[\sum_{i=1}^6 \frac 12 D_{\mu} \phi_i D^{\mu} \phi_i
  \right] + \ldots\right\}  
\end{equation}
The relevant portion of the action includes 
only kinetic terms for the four Majorana
fermions and six real scalars, and their interactions with the background
field through
the gauge covariant derivative. We have omitted in the above, 
the Yukawa couplings and the $\N=4$ quartic scalar potential.
Ultimately, integrating over the fluctuating quantum fields $\phi_i$,
$\psi_A$, the gauge fluctuations 
$A_\mu^{\rm qu}$, and the ghosts arising from gauge fixing, will 
generate an effective potential for the classical gauge fields. 

Working in Euclidean space, on ${\mathbb R^3}\times S^1$ with
antiperiodic boundary conditions for fermions around the thermal
circle, we will treat each of the quantum fluctuations separately below. 

\subsection{Gauge field fluctuations}
\label{sec:gauge_fields}
The one-loop calculations outlined in this section follow essentially
standard steps, however we include them here for  completeness.
As previously seen, the gauge field $A^\mu$ consists of
classical and quantum parts, thus the gauge part of the action can be
separated accordingly
\begin{equation}
S_A = S_A^{\rm cl}+S_A^{\rm qu}.
\end{equation}
Letting $q$ be a general function of $z$, and using the
fact that the only non-zero classical gauge field is $A_0^{\rm cl}$, the
classical action can be evaluated simply on this background,
\begin{eqnarray}
\mathcal{L}_A^{\rm cl}\, =\, \Tr\left[\frac 12 (F_{\mu\nu})^2\right] \,=\,
\Tr\left[(\partial_z A_0^{\rm cl})^2\right] =\Tr\left[\frac{4 \pi^2
    T^2}{g^2_{\rm YM} N^2}(\partial_z q)^2 Y_k^2\right].
\end{eqnarray}
Using $\Tr Y_k^2 = N k (N-k)$, we obtain 
\begin{equation}
S_A^{\rm cl}=\frac{4 \pi^2
  T^2}{g^2_{\rm YM} N}\,k(N-k)
\,\int\!\! d^3\!x \,\int_0^\beta\!\!d\tau \,(\partial_z q)^2.
\end{equation}
As mentioned briefly at the start of Section
\ref{sec:n=4_kwall}, the classical action alone is not enough to show
the existence of $k$-walls, since it is only
sensitive to the gradient energy that is minimized by a constant
$q$ solution. To find a $k$-wall solution, the action needs to be 
calculated beyond tree level, at one-loop order, self-consistently at
weak coupling. 
 
To treat the gauge field fluctuations at one-loop or quadratic order,
we shift attention to the quantum part of the action wherein we 
must fix a gauge. We employ the usual background field $R_{\xi}$ gauges
to obtain the action for the quantum fluctuations of the gauge field
\begin{equation}
\mathcal{L}_A^{\rm qu} = \Tr\left[\frac 12 (F_{\mu\nu}^{\rm qu})^2\right] +
\Tr\left[\frac 1 \xi \left(D_{\rm cl}^\mu A^{\rm qu}_\mu\right)^2\right]+
\Tr\left[\overline{\eta}\left(-D_{\rm cl}^2 \right)\eta\right] 
\end{equation}
with $\bar\eta$ and $\eta$ being the Fadeev-Popov ghosts,
and the adjoint covariant derivative $D_\mu$ and $D_\mu^{\rm cl}$ defined
thus,
\begin{equation}
D_\mu=\partial_\mu - i g_{\rm YM} [A_\mu,\,.],\,\,\,\,\,D_\mu^{\rm
  cl}=\partial_\mu - 
i g_{\rm YM} [A_\mu^{\rm cl},\,.]\,. 
\end{equation}

At the one-loop order we can completely ignore the interactions
between different quantum fluctuations. This amounts to replacing 
the full covariant derivative $D_\mu$ with $D_{\mu{\rm cl}}$ which
is gauge-covariant with respect to the background. 
Integrating by parts and assuming that the background field is
{\em constant}, the action for the quantum fluctuations becomes 
\begin{equation}
S_A^{\rm qu}=\!\!\int\!\! d^3\!x \,\int_0^\beta\!\!d\tau
\,\Tr\left[A^{\rm qu}_\mu\left(-D_{\rm cl}^2 \,g^{\mu\nu} +(1-\frac 1 \xi)
    D_{\rm cl}^\mu D_{\rm cl}^\nu\right) A_\nu ^{\rm qu}\right]+ 
\Tr\left[\overline{\eta}\left(-D_{\rm cl}^\mu D^{\mu}_{\rm
      cl}\right)\eta\right].
\end{equation}
Technically, it is important to note that we are assuming a constant
background field and therefore we will obtain an effective potential
for constant 
field configurations only.  Nevertheless we will  
employ the same effective potential to look for non-constant domain
wall profiles. Hence this really requires the profile function to be
appropriately slowly varying. 
Performing the functional integral over the gauge
fluctuations $A_\mu^{\rm qu}$, and the ghost fields, the one-loop
contribution is 
\begin{equation}
S_A^{\rm qu}=
\frac 12
\Tr\left[\ln\left(-D_{\rm cl}^2 \,g^{\mu\nu} +(1-\frac 1 \xi) D_{\rm cl}^\mu
    D_{\rm cl}^\nu\right)\right] - 
\Tr\left[\ln\left(-D_{\rm cl}^2\right)\right].
\end{equation}
The effective action can be shown to be independent of the 
gauge fixing parameter $\xi$ due to the commutativity of the
covariant derivatives for constant backgrounds. Thus, at least for
slowly varying interface profiles we are guaranteed to obtain
gauge-invariant results and, in particular, we will choose the 
Feynman gauge, $\xi=1$ so that,
\begin{equation}
S_A^{\rm qu} = 
-{1\over
  2} \Tr\ln(-D_{\rm cl}^2).
\label{eq:S_A_quant_1}
\end{equation}
As the background field $A_0^{\rm cl}(z)$ 
present in the adjoint covariant
derivative is only non-zero along the compact direction, $\tau$, it
reduces to an ordinary derivative in the transverse directions, $x$,
$y$ and $z$. In the compact direction the background field is
proportional to the matrix $Y_k$, and being diagonal with $N$
elements, there exist non-trivial contributions to the covariant
derivative when acting upon ladder generators $t_{ij}$ and $t_{ji}$
(see for example 
Eq.(\ref{eq:gen_ladder_commutators})).  Following the notation
of \cite{korthalsaltes}
\begin{eqnarray}
&&D_0^{\rm cl} \,t_{ij}=(\partial_0 - 2\pi i T q)\,t_{ij} \equiv D_0^+
\,t_{ij}\,,\\\nonumber\\
&&D_0^{\rm cl} \,t_{ji}=(\partial_0 + 2\pi i T q)\,t_{ji} \equiv D_0^-
\,t_{ji}\,.
\label{eq:covariant_q}
\end{eqnarray}
The commutator of $Y_k$ with the ladder operators for $SU(N)$, 
has very similar properties
to the commutator in Eq.(\ref{eq:gen_ladder_commutators}) while 
there are significant differences:
\EQ
{
[Y_k, t_{ij}] = N t_{ij}\,\qquad[Y_k, t_{ji}] =- N t_{ji}\,,\qquad
i\in[N-k,N]\,,\qquad j\in[1, N-k]\label{ladderyk}
}
All other commutators vanish. 
The full non-trivial $q$ dependence comes from the
action of the  covariant derivatives on the ladder generators;
equivalently, from integrating out all off-diagonal fluctuations that
do not commute with $Y_k$. For this
reason we may replace the covariant derivatives with 
\begin{equation}
D_{\rm cl}\rightarrow(D_0^\pm, \vec{ \partial}\,).
\end{equation}
Fourier transforming to Euclidean momentum space, the
temporal derivative $\partial_0$ may be replaced by the Matsubara
frequencies $p_0$,
\begin{equation}
i \partial_0 \rightarrow p_0 = 2 \pi n T\,,\qquad n\in {\mathbb Z}.
\end{equation}  
The action of the covariant derivatives on ladder operator-like
fluctuations is, from Eq.(\ref{eq:covariant_q}), 
\begin{equation}
iD_0^\pm \rightarrow p_0^\pm = 2 \pi T (n \pm q).
\label{eq:co-var_der_1}
\end{equation}

Since there are precisely $k(N-k)$ fluctuations \eqref{ladderyk} which
yield a non-zero contribution to the effective action at one-loop 
and the sum over the Matsubara modes includes both positive
and negative integers, we have 
\begin{equation}
S_A^{\rm qu} = \,2k(N-k)\,{V}_{\rm tr} L 
\sum_{n=-\infty}^{+\infty} \int T 
\frac{d^3\textbf{p}}{(2\pi)^3}\ln\left((p_0^+)^2+\textbf{p}^2\right). 
\end{equation}
Here $V_{\rm \tr}$ is the volume of the space transverse to the $z$-axis
\EQ{
V_{\rm tr}=L_1 L_2\beta\,,
} 
with $L_1, L_2$ being the extents of the system in the $x$ and $y$ 
directions respectively. As the Euclidean 
time circle is compactified, the $k$-wall
can be viewed as being smeared along this direction.
Now it remains to determine the dependence of the one-loop effective
action on $q$. Up to irrelevant additive constants, the $q$-dependence
is determined by the variation of $S_A^{\rm qu}$ with respect to $q$,
\begin{eqnarray}
\frac{1}{2\pi T}\,\frac{\partial S_A^{\rm qu}}{\partial q} &=&\,4k(N-k)
V_{\rm tr} \,L \sum_{n=-\infty}^{+\infty} \int T \,\frac
{d^3\textbf{p}}{(2\pi)^3}\left(\frac{p_0^+}{(p_0^+)^2+\textbf{p}^2}\right)\,
\\\nonumber
&=&\,-4k(N-k)V_{\rm tr}\,L \,\pi T^3 \sum_{n=-\infty}^{+\infty}
(n+q)|n+q|,
\end{eqnarray}
where the final result follows from 
standard expressions for the regulated integral using dimensional
regularisation.
Using zeta function regularisation, the explicitly
divergent sum over $n$ can be controlled quite elegantly
\footnote{The Hurwitz zeta function is defined as 
$\zeta(l,m)=\sum_{n=0}^{+\infty}(n+m)^{-l}$.},
\begin{equation}
\sum_{n=-\infty}^{+\infty} (n+q)|n+q|= 
\sum_{n=0}^{+\infty}\left[(n+q)^2 - (n+1-q)^2\right] = 
\zeta(-2, q) - \zeta(-2,1-q).
\end{equation}
Remarkably, this 
particular form of the Hurwitz zeta function is a simple polynomial in $q$ 
\begin{equation}
\zeta(-2,q)= -\frac 1{12} \frac d{dq}\left[ q^2(1-q)^2\right].
\label{eq:zeta_der}
\end{equation}
Finally we obtain the one-loop effective potential for slowly varying 
$q(z)$, (up to an additive constant)
\begin{equation}
S_A^{\rm qu} = \,\frac 43 k(N-k) V_{\rm tr}\,\pi^2T^4\,\int_0^L dz \,q^2(1-q)^2.
\end{equation}
The potential is manifestly invariant under $q\rightarrow 1-q$ and has
minima at $q=0$ and $q=1$. We can therefore have a `kink'-like
configuration interpolating between these two vacua. Combining the 
quantum one-loop action along with the classical kinetic term
calculated earlier, the total effective action can be compactly
expressed, after 
a coordinate rescaling $z\rightarrow z^\prime= \sqrt{g^2_{\rm
    YM}N/3}\,T z$ as,
\begin{equation}
S_A = \frac {4\pi^2 T^3}{\sqrt{3N} g_{\rm YM}}\,k(N-k)\,V_{\rm tr} 
\int_0^{L'} dz^{\prime}\,
\left[ \left(\frac{\partial q}{\partial z^\prime}\right)^2 +
  q^2(1-q)^2\right].
\end{equation}
The double well potential, represents the so-called
``$q$-valley". Due to the rescaling of $z\rightarrow z^\prime$, the
upper limit of integration $L$ is
also effectively rescaled:  $L\to L^\prime=
 \sqrt{g^2_{\rm YM} N/3}\,TL$. 
The large volume limit corresponds to
 $ \sqrt{g^2_{\rm YM} N} TL\to \infty$ which can also be viewed as the 
three dimensional limit when the thermal circle shrinks to zero size. 

We can now self-consistently justify the use of the constant-$q$
effective potential to infer the existence of the spatially varying
domain wall. In terms of the $z'$ coordinate, it is clear that the
width of the domain wall is a number $\sim {\cal O}(1)$. In physical
units, the width of 
the domain wall is then set by $(\sqrt{g^2_{\rm YM} N} T)^{-1}$, which
is the Debye or electric screening length. At weak gauge coupling (or
weak 't Hooft coupling at large $N$), this is much larger than the
typical thermal wavelength $T^{-1}$, of the perturbative degrees of
freedom. Thus the domain wall is thick and a slowly
varying configuration. 
Furthermore, since the scale of variation is set by the Debye
scale, the wall and its properties should be accessible in
perturbation theory. If the wall thickness had been set by the
magnetic scale $(g^2_{\rm YM} N T)^{-1}$, the non-perturbative scale of
the three dimensional effective theory, the perturbative description
above would be invalidated.

Having reviewed the perturbative gauge field contributions to the
physics of the domain walls, let us now turn to the matter fields in
the adjoint representation in the $\N=4$ theory.

\subsection{Scalar field fluctuations}
The $\N =4$ theory has six hermitian scalars transforming in the
adjoint representation of the gauge group. We consider them first, due 
to the similarity in
the calculation to the gauge field contribution, 
before turning to the fermion fields in Section
\ref{sec:n=4_fermion}. The scalar part of the action coupled to the
classical background gauge field is, 
\begin{equation}
S_S = \int d^3x \int_0^\beta d\tau \,\Tr\left[\sum_{i=1}^{n_s}\frac 12
  \left(D_\mu^{\rm cl}\phi_i D^\mu_{\rm cl}\phi_i \right)\right] 
\end{equation}
where $n_s=6$ is the number of real adjoint scalars, and we have ignored
interactions of the fluctuations.
Integrating out the scalar field fluctuations we have
\begin{equation}
S_S = \frac{n_s}{2} \!\!\int\!\! d^3\!x \,\int_0^\beta\!\!d\tau\,
\Tr\ln(-D_{\rm cl}^2).
\end{equation} 
This is exactly the same as Eq.(\ref{eq:S_A_quant_1})
up to the overall normalization and generates the one-loop potential
\begin{equation}
S_S=\frac {4\pi^2T^3}{\sqrt{3N g_{\rm YM}^2}}\,
k(N-k) \frac {n_s}2 V_{\rm tr}\int_0^{L'} dz^\prime \,q^2(1-q)^2. 
\end{equation}

\subsection{Fermionic contributions at one loop}
\label{sec:n=4_fermion}

Finally there are the $n_f=4$ fermions transforming as a 
${\bf 4}$ of the $SO(6)$ R-symmetry. These play a crucial role at
finite temperature since they have antiperiodic, supersymmetry
breaking boundary
conditions around the thermal circle. At quadratic order in the
fluctuations 
\begin{equation}
S_F = \!\!\int\!\! d^3\!x \,\int_0^\beta\!\!d\tau \,\Tr\left[
  \sum_{A=1}^{n_f} \overline{\psi}_A D\!\!\!\!\slash\,\,\,\psi_A\right]. 
\end{equation}
Working in Euclidean
space, the Dirac gamma matrices are 
\begin{equation}
\gamma^{1,2,3} = \left(
\begin{array}{cc}
0&- i \sigma_{1,2,3}\\
i \sigma_{1,2,3}&0
\end{array}
\right),\,\,\,\,\,
\gamma^4 = \left(
\begin{array}{cc}
0&{\bf 1}\\
{\bf 1 }&0
\end{array}
\right)
\end{equation}
 where $\sigma^i$ are the standard Pauli matrices.
The functional integral over the fermion fields then yields the
Pfaffian of the Dirac operator, since $\psi_A$ and $\bar \psi_A$ are
not independent due to the Majorana condition,
\begin{eqnarray}
S_F
=-\!\!\int\!\! d^3\!x \,\int_0^\beta\!\!d\tau \,n_f\,
  \Tr\ln\left[-((D_0^{\rm cl})^2 + {\bf \nabla}^2)\right].
\label{fermidet}
\end{eqnarray}
Despite the formal similarity to the bosonic contributions, it is at
this point that the analysis departs from that for scalar and gauge
fluctuations.

Consider first the case (zero temperature) 
wherein the compact direction has  {\em periodic} (SUSY
preserving) boundary
conditions for the fermions. The fluctuation determinant
\eqref{fermidet} would be identical to that of the bosons (and the
opposite sign) to produce a one-loop action of the form 
$S_F = - n_f S_A^{\rm qu}$. With supersymmetric boundary conditions
the three fluctuation terms at the one-loop level 
would cancel, leaving only the classical action, 
\begin{equation}
S_{\rm Total} = S_A^{\rm cl} +S_A^{\rm qu} +S_S+S_F = S_A^{\rm cl} +(1+n_s/2 -
n_f)S_A^{\rm qu} = S_A^{\rm cl}. 
\end{equation}
The cancellation between bosons and fermions will persist at all loop
orders for SUSY-preserving boundary conditions.

However, in the Euclidean thermal theory, 
the fermions have anti-periodic boundary
conditions around the Euclidean time circle. The Matsubara 
frequencies are thus shifted to half-integer values $n\to n+\frac 12$,
$n\in {\mathbb Z}$.  The eigenvalues of the covariant derivative, 
Eq. (\ref{eq:co-var_der_1}), acting on fermions is modified 
\begin{equation}
iD_0^\pm \rightarrow p_0^\pm = 2 \pi T \left(n+\frac 12 \pm q\right).
\end{equation}
 This shift has a non-trivial effect on the one-loop effective
 potential and the fermi-bose cancellations will be absent. 
With a half-integer moding of the fermionic Matsubara frequencies, 
the sums over $n$ for $p_0^+$ and
$p_0^-$ are no longer equivalent, and each sum must be
evaluated separately,
\begin{equation}
S_F = - k(N-k) V_{\rm tr} L T \,n_f \sum_{n=-\infty}^{+\infty} 
\int\frac {d^3\textbf{p}}{(2\pi)^3}
\left( 
\ln\left[(p_0^+)^2+\textbf{p}^2\right]
+\ln\left[(p_0^-)^2+\textbf{p}^2\right]
\right).
\end{equation}
Once again as before it is useful to 
take the variation of the action with $q$, in order to determine the 
$q$-dependence of the effective potential,  
\begin{equation}
\frac{\partial S_F}{\partial q} = -2k(N-k) (2\pi T) V_{\rm tr} L T \,n_f
\sum_{n=-\infty}^{+\infty} \int\frac {d^3\textbf{p}}{(2\pi)^3} 
\left(
\frac{p_0^+}{(p_0^+)^2+\textbf{p}^2}-\frac{p_0^-}{(p_0^-)^2+\textbf{p}^2}
\right).
\end{equation}
Again, integrating over the spatial momenta $\textbf{p}$
employing dimensional regularization, 
\SP
{
&\frac{\partial S_F}{\partial q} = 2k(N-k) (2\pi T)V_{\rm tr} L T \,n_f\,
\pi T^2 \\ 
&\times\sum_{n=-\infty}^{+\infty}
\left[
\left(n+1/2+q\right)\left|n+1/2+q\right| -
\left(n+1/2-q\right)\left|n+1/2-q\right|  
\right].  
}
Now, to regulate this sum with zeta functions, we consider the sum in two separate regions of $q$, $q \in [0,1/2]$  \& $q\in[1/2,1]$:
\begin{eqnarray}
q \in [0,1/2] &\rightarrow& 2 \sum_{n=0}^{+\infty}\left[ \left(n+1/2+q\right)^2 - \left(n+1/2-q\right)^2\right]\\
q \in [1/2,1] &\rightarrow& 2 \sum_{n=0}^{+\infty}\left[ \left(n-1/2+q\right)^2 - \left(n+3/2-q\right)^2\right]
\end{eqnarray}

As applying the shift $q\rightarrow 1-q$ swaps the two terms, and their respective regions of $q$, there is no loss of generality to simply consider the region $0\leq q \leq 1/2$, and introduce a overall doubling factor. The definition of the Hurwitz zeta function as a derivative,
Eq.(\ref{eq:zeta_der}) allows the action to be explicitly determined,
up to integration constants, 
\SP
{
&S_F = -\frac{4\pi^2T^3}{\sqrt{3N}g_{\rm YM}} k(N-k) V_{\rm tr}\,n_f\times 2 \int
dz\left(\frac 12+q\right)^2\left(\frac 12-q\right)^2. 
}
The integration region is now only defined over $0\leq q\leq 1/2$. It is obvious now that this fermionic action will not cancel against the quantum gauge and scalar parts.

\subsection{Full one-loop effective action}
Putting all the above ingredients together we obtain the full one-loop
effective action for the interface, and adjusting the integration region accordingly;
\begin{eqnarray}
S_{\rm Total} &=& S_A+S_S+S_F\\
&=&\frac {4\pi^2T^3}{\sqrt{3N} g^2_{\rm YM}}\,k(N-k)\,V_{\rm tr} \,2 \int
dz^{\prime} \nonumber\\ &&\times\, 
\left[ \left(\frac{\partial q}{\partial z^\prime}\right)^2 +
  (1+\frac{n_s}{2})
  q^2(1-q)^2 
- n_f  \left(\frac 12+q\right)^2\left(\frac
      12-q\right)^2 \right].
\end{eqnarray}
Letting $n_s$ and $n_f$ take their explicit values in $\N=4$ SYM, 
the quantum effective action simplifies to 
\begin{equation}
S_{\rm Total}=\frac {4\pi^2T^3}{\sqrt{3 g^2_{\rm YM}N} }k(N-k) V_{\rm tr} \, \times 2
\int dz^{\prime} \left[ \left(\frac{\partial q}{\partial
      z^{\prime}}\right)^2+ 2q^2(3-4q) -1/4\right].
\end{equation}
It is a simple exercise to obtain the minimum action configuration
that interpolates between the two vacua $q=0$ and $q=1/2$, satisfying 
$(dq/dz')^2 = 2q^2(3-4q)$, so that the action for the kink or domain
wall is 
\begin{eqnarray}
S_{\rm Total}&=& \frac {4\pi^2T^3}{\sqrt{3g^2_{\rm YM} N} }k(N-k) V_{\rm tr} \times\,4\int_0^{1/2} dq \sqrt{2q^2(3-4q)}\\
&=& \frac {4\pi^2T^3}{\sqrt{3g^2_{\rm YM} N} }k(N-k) V_{\rm tr} \frac{\sqrt 2}{5} (3\sqrt 3 -2)
\end{eqnarray}

We therefore conclude that the tension of the $k$-wall in the
Euclidean high temperature, $\N=4$ theory at weak coupling is\footnote{We would like to thank C. Korthals-Altes for pointing out an error in a previous archive version of the paper.} 
\begin{eqnarray}
{\cal T}_k &=& \frac {4\pi^2T^2}{\sqrt{3g^2_{\rm YM} N} }k(N-k) \frac{\sqrt 2}{5} (3\sqrt 3 -2)\\
&\approx&0.904 \frac {4\pi^2T^2}{\sqrt{3g^2_{\rm YM} N} }k(N-k)
\label{weaktension}
\end{eqnarray}

where one factor of $T$ has cancelled against the size of thermal circle in 
$V_{\rm tr}$, leaving us with the tension of a $1+1$ dimensional
interface in three dimensions.
We note firstly that the parametric dependence on the gauge coupling 
is the same as in ordinary Yang-Mills theory 
\cite{Bhattacharya:1990hk,
  Bhattacharya:1992qb,korthalsaltes}. One difference is that unlike in
pure Yang-Mills theory, the gauge coupling itself does not run and
therefore does not depend on temperature. We are, however, free to
choose an arbitrary weak coupling in $\N=4$ theory, $g_{\rm YM} \ll
1$. All other qualitative aspects of the solution are similar to pure
Yang-Mills theory. Specifically, the wall is ``fat'' with a width set
by the Debye screening length $(\sqrt{g^2_{\rm YM} N}
T)^{-1}$. Interestingly, for $k\sim {\cal O}(N^0)$, in the 't Hooft
large-$N$ limit, the $k$-wall tension scales as $N^1$ rather than $N^2$.

Finally, the one-loop calculation demonstrates a Casimir scaling law for
the tension of the $k$-wall. It is not {\it a priori} clear that
Casimir scaling will persist at higher loop orders, since at one-loop
its origin is essentially kinematic.  
Next we will investigate  
whether Casimir scaling remains at the two-loop  level in $\N=4$ theory.  
\section{\nfour at 2-loop}
It has been shown \cite{Giovannangeli:2001bh,Giovannangeli:2002uv}
in pure Yang-Mills theory that Casimir scaling of ${\mathbb Z}_N$ 
domain walls remains at two-loops, but is lost at three-loops. 
Below we
adapt the arguments of the two-loop result for pure Yang-Mills
theory to argue that the scaling for $\N=4$ theory will also be
Casimir like at the two-loop order. 
\subsection{Pure YM at 2-loop}
Consider first the 2-loop calculation of the domain wall tension in 
pure Yang-Mills theory in the deconfined phase. Let us first lay out
our notation and conventions. Defining the structure constants of the
$SU(N)$  algebra and their normalisation as usual
\EQ
{i f^{a,b,c} = 2 \,\Tr \left(\left[t^a,t^b\right]t^c\right)\,,\qquad
\left(f^{a,b,c}\right)^2 = \frac 12\,,}
the indices $a$, $b$, and  $c$ can stand for Cartan generators, 
$t_{\rm diag}$, or the ladder generators $t_{ij},\,t_{ji}$. From the
commutation relations for the ladder generators and the Cartan
generators, it follows that the only non-zero values of $f^{a,b,c}$ 
exist when no more than one of the generators is diagonal. 
The specific non-zero cases are explored in more detail below. 

It was demonstrated in \cite{Giovannangeli:2001bh,
Giovannangeli:2002uv, korthalsaltes} that at the
 two-loop level, including three
 and four vertex gluon interactions, and gluon-ghost
 interactions, all possible loop graphs generate contributions to the
 $k$-wall action, of the form 
\begin{equation}
S_2 \sim \sum_{a,b,c} f^{a,b,c}f^{a, b,c} B_2(C_a) B_2(C_b).
\label{eq:2loop_structure_expression}
\end{equation}
Here $B_2$ is the second Bernoulli polynomial
 which is even in $C_a$ 
\begin{equation}
B_2(C_a) \sim \left(C_a^2 - |C_a|+{1\over 6}\right)
\end{equation}
 and the variables $C_a$ are shorthand for the functions $C_{ij}$ that
encode all the $q$ dependence   
\begin{equation}
C_{ij}= A_{0\,i} - A_{0\,j} \sim q \left[(Y_k)_i-(Y_k)_j\right].
\end{equation}
 It is obvious that $C_{ii}=0$, while
 $C_{ij}$ is only non-zero when $i$ and $j$ sit in different ``sectors'' of
 $Y_k$ (recall from \eqref{hypercharge} that its elements live in two
 sectors taking only two possible values). Thus $C_{ij}= 0$
 or $\pm q$ up to an overall  factor.  

 With the above definitions and conventions, explicit computation of 
Eq. (\ref{eq:2loop_structure_expression})
reveals Casimir-like scaling, like that at one-loop. This arises from
summing all non-vanishing terms in
\eqref{eq:2loop_structure_expression} that have $q$
dependence ({\it i.e.} ignoring all terms proportional to 
 $B_2(0)^2$). The different non-trivial terms can be classified and
 accounted for as explicitly explained in the Appendix. The final
 result of the combinatorics gives  
\EQ
{
S_2 \sim Nk(N-k)\left[B_2(q)^2+2B_2(q)B_2(0)\right].
}
As explained in the Appendix, the two key technical reasons for Casimir scaling
in the final result are: one, that all $q$ dependence in
\eqref{eq:2loop_structure_expression} arises from terms where at least
one of the two indices $a$ and $b$ are off-diagonal
generators. Secondly, and perhaps more importantly, the combination 
$B_2(C_a) B_2(C_b)$ is an even function of the $C_a$.

\subsection{Argument for Casimir scaling in $\N=4$ SYM at 2-loops} 
The two factors outlined in the previous section, coupled with the 
structure constants,  effectively guarantee Casimir
 scaling in pure Yang-Mills. For this to be present in $\N=4$ theory,
 the same  factors must come into play.  
The propagators for the adjoint scalars in 
 $\N=4$ SYM, are equivalent (up to an overall factor) to the ghost
 propagators in pure YM. Therefore we expect the inclusion of the
 adjoint scalars, to not change the Casimir scaling at two loops. 

For the adjoint fermions of the $\N =4$ theory, a SUSY-based 
argument can be employed. With periodic boundary conditions, all
perturbative fermionic and bosonic contributions to the effective
potential for a constant 
(slowly varying) $A_0$ background will cancel due to supersymmetry. 
Since the bosonic fluctuations 
at two-loop yield Casimir scaling of the effective potential, 
the fermionic contributions will exhibit the same. 

 When the boundary conditions on the fermions are changed so that we
 have a thermal interpretation, the anti-periodic boundary conditions on
 fermions  will only lead to a shift in $q$ dependence, $q\rightarrow
 q^\prime = q\pm 1/2$ due to the change in the Matsubara modes. 
This shift would leave all other overall scaling factors intact and we expect 
\begin{equation}
S_2^F \sim (f^{ij,ji,\,{\rm diag}})^2 B_2(C_{ij}^F) B_2(C_{ji}) + 
\mathrm{permutations}
\end{equation} 
where $C_{ij}^F$ is the shifted difference,
\begin{equation}
C_{ij}^F \sim (A_{0\,i} - A_{\,0j}) \pm \frac 12 \sim q\pm \frac 12.
\end{equation}
Such a term would arise from the two-loop graph shown in Figure 1. 
\begin{figure}[h]
\begin{center}
\epsfig{file=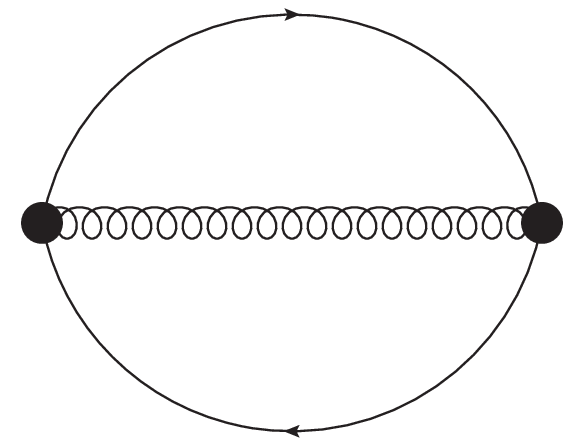, width =1.8in}
\end{center}
\caption{\footnotesize A two-loop fermion contribution to the
  effective action.}
\end{figure}

 This would then imply 
\begin{eqnarray}
S_2^F &\sim& \frac 12 B_2(q) \left[ B_2\left(q+\frac 12\right) + B_2
  \left(q-\frac 12\right)\right]k(N-k) \\\nonumber\\\nonumber
&=& \frac 12 q^2 (1-q)^2 \left[
  \left(\frac 12+q\right)^2 \left(\frac 12-q\right)^2 +\left(-\frac
    12+q\right)^2\left(\frac 32-q\right)^2\right] k(N-k)
\end{eqnarray} 
representing the pair of gluon-fermion vertices in Figure 1. These
arguments make it plausible that Casimir scaling of $k$-wall tensions
persists at the two-loop level in $\N=4$ SYM. However, there is no
reason why this should continue to be the case beyond two-loops

\section{${\mathbb Z}_N$ domain walls at strong coupling}
\label{strong-coupling}

We now turn to the domain walls in strongly coupled ${\cal N}=4$
theory in the large $N$ limit, at finite temperature. The 
deconfined phase of the four dimensional field 
theory at strong coupling is described by Type IIB string theory in
the Schwarzschild black
hole in $AdS_5\times S^5$ \cite{witten}. In the Euclidean picture, the conformal
boundary of the geometry is ${\mathbb R}^3\times S^1$, the boundary
$S^1$ being identified with the 
Euclidean thermal circle of the strongly coupled field  theory.
The spontaneous breaking of the 
${\mathbb Z}_N$ center symmetry in the deconfined phase follows from
the fact that the thermal circle shrinks smoothly to zero size at a
radial coordinate corresponding to the horizon of the Lorentzian black hole.  
The resulting black hole cigar can be wrapped by finite action string
world-sheet instantons resulting in a non-zero Polyakov loop that
spontaneously breaks the ${\mathbb Z}_N$ symmetry.

\subsection{D-string as domain wall}
With the ${\mathbb Z}_N$ center symmetry spontaneously broken it
should be possible to identify the $N$ distinct ``vacua'' of the
theory in the IIB string dual. This question was raised and addressed
in \cite{Aharony:1998qu}. Shifting the phase of the Polyakov loop
by the $N$th roots of unity, moves us through  
the $N$ possible ground states of the deconfined theory. In
the string dual this is realized as a shift of the 
NS two-form field $B^{(2)}_{NS}$ integrated over the black
hole cigar $D_2$,
\EQ
{
\int_{D_2} B_{NS} \rightarrow\int_{D_2} B_{NS} +2\pi k/N\,,
\qquad\,k=1,\ldots N.
\label{shift}
}
Disc instantons wrapping the black hole cigar will
then pick up a phase $\exp(i \,2\pi k/N)$, determining the VEV of the
Polyakov loop \footnote{The $N$ vacua should be physically
  equivalent. This can be understood from the transformation of the 
term in the Type IIB low energy effective action 
$S = i\int B_{NS} \wedge {H_{RR}\over 2\pi}
\wedge {F_5\over 2\pi}$. In the sector with 
$\int_{{\mathbb R}^3} H_{RR}/(2\pi)=1$, the partition function will be
invariant under \eqref{shift}, as 
$\delta S= i \frac{2\pi k}{N} \,N\,\int_{{\mathbb R}^3}
{H_{RR}\over 2\pi} = 2\pi i k$,
where we have used the fact that there are $N$ units of five-form flux
through the $S^5$.}.

Across a domain wall, the phase of the Polyakov loop jumps. The
argument of \cite{Aharony:1998qu}, shows that a D-string
world-sheet $\Sigma \subset {\mathbb R}^3$ (and pointlike on the disc $D_2$
and the $S^5$) provides precisely such a jump. In particular, across
$\Sigma$, the RR three form flux $H_{RR}$ changes by one unit for a
single D-string. 

There is another argument establishing the connection between
${\mathbb Z}_N$ domain walls and the D-string. This exploits the
direct relation between ${\mathbb Z}_N$ interfaces and spatial 
't Hooft loops \cite{KorthalsAltes:1999xb, KorthalsAltes:2000gs}. 
The spatial 't Hooft loop operator $V(C)$, 
along a contour $C$, creates an infinitely thin
tube of chromomagnetic flux along $C$. The spatial 't Hooft loop bounds a
surface (a Dirac sheet) across which the gauge potential $A_0$ is
discontinuous. The explicit perturbative  computation of the
expectation value of the  't Hooft loop in the deconfined phase can be
shown to reduce to the domain wall calculations presented earlier in
this paper. Specifically, the leading contribution to a large 
't Hooft loop in ${\mathbb R}^3$ 
is proportional to the area $A$ of the minimal surface bounded by $C$,
$V(C)\sim \exp(-{\cal T} \,A)$. Thus the 
infinite volume domain wall tension ${\cal T}$ is computed by the spatial 't
Hooft loop of infinite extent. 

By the AdS/CFT correspondence, the 't Hooft loop is a Euclidean
D-string world-sheet with disc topology, whose boundary traces the
spatial loop in the field theory on the conformal boundary of the
spacetime. In the AdS-Schwarzschild black hole geometry, the D-string 
world-sheet droops toward the interior of the spacetime. As the size
of the loop is scaled up, most of the D-string worldsheet sits at the
horizon where the bulk Euclidean geometry smoothly ends, and
consequently the spatial 't Hooft loop exhibits an area law
\cite{Witten:1998zw, Gross:1998gk, Brandhuber:1998er}.  
When the loop is taken to be of infinite extent, we obtain a 
Euclidean D-string world-sheet $\Sigma \subset {\mathbb R}^3$
located at the horizon of the Euclidean black hole. This is the
${\mathbb Z}_N$ interface.

\subsection{$k$-wall tensions}

The metric for the high temperature, Euclidean AdS-Schwarzschild black
hole in $AdS_5\times S^5$ is,
\SP
{
ds^2= & R^2 \left[f(r)\,dt^2+ {dr^2\over f(r)}+ r^2\,d\vec x
  ^{\,2}+ d\Omega_5^2\right]\,\label{metric}\\\\ 
f(r) = & \,r^2- {\pi^4 T^4\over r^2}.
}
where $R^4 = 4\pi (g_{s} N) \alpha^{\prime 2} = (g^2_{\rm YM}
N)\alpha^{\prime 2}$. The D1-brane
world-sheet theory is described by the Dirac-Born-Infeld action in the
absence of background $C^{(2)}_{RR}$ potential
\EQ
{
S_{D1} = {1\over 2\pi\alpha'}\int d^2\sigma\, e^{-\Phi}\,\sqrt{{\rm det}
  {}^* g}.
}
The dilaton is constant with $e^{-\Phi}= 1/g_s = 1/g^2_{\rm YM}$.
Assuming that the D-string worldsheet $\Sigma$ is oriented along the 
$x-y$ plane, we choose the embedding $\sigma_1=x$ and $\sigma_2=y$. 
Then 
\EQ
{
{\rm det}{}^{*}g = R^4 r^4.
}
and to minimize the action, the D-string will sit at the smallest
possible value of $r$, which, in this geometry is $r=\pi T$. We then
find the tension of the $k=1$ wall at strong coupling is 
\EQ
{
{\cal T}_1 = {1\over 2\pi \alpha' g_s} R^2 \pi^2 T^2 = 2\pi^{2}\,
{N\over \sqrt{g^2_{\rm YM}N}}\,T^2.
}
Remarkably, the parametric dependence of this formula, on the 't Hooft
coupling and $N$, closely resembles \eqref{weaktension}. The
dependence on the temperature is guaranteed to be quadratic by the
underlying conformal invariance of the $\N=4$ theory. The $N$
dependence is consistent with the domain wall being a D-brane in the
large-$N$ limit and the fact that the tension of the D-string in AdS is
proportional to $1/\sqrt{g^2_{\rm YM} N}$ is also obvious from
supergravity. What is interesting to note is that the formula at weak coupling
also has the same dependence on the 't Hooft coupling.

For a collection of $k$ D-strings, with $k\sim {\cal O}(1)$, the
tension is simply $k$ times that of a single D1-brane in the
AdS-Schwarzschild background.

When the number of D-strings $k$, becomes of order $N$, in the large
$N$ limit, we can no longer think of the system as consisting of $k$
separate D1-branes. In fact we expect the collection to blow up into a
higher dimensional brane via an analogue of the dielectric effect
\cite{Myers:1999ps} in the curved geometry. There are two possible
blown up brane configurations to consider in the $AdS_5\times S^5$
black hole geometry, carrying $k$ units of D-string charge. At zero
temperature, in $AdS_5\times S^5$, electric Wilson loops in the
$k^{\rm th}$ rank antisymmetric and symmetric tensor representations
of $SU(N)$, are computed by a D5-brane wrapped on an $S^4\subset S^5$
and a D3-brane wrapping an $S^2\subset AdS_5$, respectively 
\cite{Drukker:2005kx, Hartnoll:2006hr, yamaguchi, Hartnoll:2006is}. 
Hence a collection of $k$ D-strings, representing 't Hooft loops, 
could expand into wrapped NS5 and D3-branes, by S-duality.

\subsubsection{The $k$-wall as a 5-brane}

We expect that the correct configuration describing a $k$-wall is
an expanded 5-brane. The 5-brane yields the $k^{\rm th}$ rank 
antisymmetric tensor representation of Wilson/'t Hooft loops; this is 
manifestly symmetric under $k\rightarrow N-k$, a property that we
require from a candidate ${\mathbb Z}_N$ interface 
\footnote{'t Hooft/Wilson loops in the symmetric tensor representation do not
  have this symmetry property.}. The intimate relationship between
the baryon vertex (a 5-brane in the bulk) 
and flux tubes in the gauge theory 
(D- and F-strings in the bulk) 
\cite{wittenbaryon,calguij1,calguij2}
also naturally leads us to consider
5-branes as the candidates.

It is most convenient to first study an expanded probe D5-brane carrying
$k$ units of F1-string charge, and subsequently S-dualize to obtain
the D-string domain wall. The action for the probe D5-brane has both
Dirac-Born-Infeld and Wess-Zumino terms, and our analysis follows
closely that in \cite{ Hartnoll:2006hr}
\AL
{
&S= {1\over (2\pi)^5\alpha^{\prime 3}g_s}\left[
\int dx\,dy d\Omega_4 \sqrt{{\rm det}({}^* g +2\pi\alpha' F)}
- i g_s \int 2\pi\alpha'F\wedge {}^* C_4
\right]\\\nonumber\\
& C^{(4)} = {R^4\over g_s}
\left[\frac{3}{2}(\alpha-\pi)-\sin^3\alpha \cos\alpha 
-\frac{3}{2}\cos\alpha\sin\alpha
\right]\,{\rm Vol}(S^4).
}
Here, $C^{(4)}$ is the relevant component of the RR four-form
potential, proportional to the volume form on $S^4$. We have
assumed that the D5-brane wraps an $S^4$ located at a polar angle
$0 < \alpha \leq \pi$ inside the $S^5$. Thus the D5-brane has
world-volume $\Sigma\times S^4$ where $\Sigma\subset {\mathbb R}^3$ is
oriented in the $x-y$ plane. In addition, a world-volume electric
field $F_{xy}$ in the $x-y$ plane is switched on to endow the wrapped
5-brane with F-string charge. Since we are working in Euclidean
signature, the electric field is imaginary, so it is useful to define
$F_{xy}=i F$.

With this ansatz, using $R^4= 4\pi (g_s N)\alpha^{\prime 2}$,
the D5-brane action is 
\EQ
{
S = {N\sqrt{\lambda}\over 3\pi^2}
\int dx\,dy\left(\sin^4\alpha\sqrt{r^4- {4\pi^2 F^2\over\lambda}}
-D(\alpha){2\pi F\over\sqrt\lambda}\right)
}
where we have defined the 't Hooft coupling $\lambda=g^2_{\rm YM}N$
and
\EQ
{
D(\alpha)=-\frac{3}{2}(\alpha-\pi)+\sin^3\alpha \cos\alpha 
+\frac{3}{2}\cos\alpha\sin\alpha.
}
The equation of motion for the gauge field associated to 
$F$ gives the total F-string charge $k$ which is quantized; 
in particular the
canonical momentum $\delta S/\delta F$ is the coupling of the
world-sheet to the $B_{NS}$ field. Thus,
\EQ
{
k= - {\delta S\over \delta F}= {2N\over 3\pi}
\left({2\pi F\over\sqrt\lambda} 
{\sin^4\alpha\over\sqrt{r^4 - \frac{4\pi^2 F^2}{\lambda}}}+D\right).
\label{eomf}
}
Together with this, the equation of motion for the polar angle
$\alpha$ determines the angle (and the size of the $S^4$) completely
in terms of the string charge $k$, 
\AL
{
&{{2\pi F/ \sqrt\lambda}\over\sqrt{r^4 - \frac{4\pi^2 F^2}{\lambda}}} =
-\cot\alpha\label{f}\\
&\cos\alpha\sin\alpha -(\alpha-\pi) = {k\over N}\pi.
\label{anglek}
}
This equation implicitly fixes the location of the $S^4$ inside
the $S^5$. Importantly, under the operation $k\rightarrow N-k$, the
associated polar angle $\alpha$ is mapped to $\pi-\alpha$. All
physical properties of the wrapped object are therefore invariant
under $k\rightarrow N -k$, as necessary for the (magnetic) ${\mathbb Z}_N$ interface.

Since the world-volume electric field and the size of the internal
$S^4$ is completely determined, it only remains to verify the radial
coordinate of the F-string configuration. Plugging the solutions
\eqref{f}, \eqref{anglek}, we have the effective action
\EQ
{
S = {N\sqrt\lambda\over 3\pi^2}\int dx\,dy \, r^2 \left[\sin^3\alpha +
{3\over 2} \left({k\over N} \pi\right)\cos\alpha \right].
}
\subsubsection{Boundary terms}
An extremely important point here is the inclusion of ``boundary
terms'' in the problem at hand, an issue which was tackled in 
\cite{ Hartnoll:2006hr} and related references cited above. 
In these latter works, Wilson loops were being computed
and the wrapped probe branes also had boundary terms in their action
that were crucial and necessary. The interface under investigation
here does not appear to have obvious boundary terms that need
to be added since the entire world volume $\Sigma \subset {\mathbb
  R^3}$ does not actually extend to the boundary of $AdS$
space. However, there is one type of boundary term that needs to be
accounted for. This term acts as a Legendre transform,
trading the world-volume gauge potential for its conjugate momentum
and fixing the string charge $k$,
\EQ
{
S_{\rm bdry}=  k \int dx \,dy \,  F .
}
so that the net action 
\EQ
{
S+ S_{\rm bdry}= 
N {\sqrt\lambda\over 3\pi^2} \int dx \,dy \, \left(r^2 \,\left[\sin^3\alpha +
{3\over 2} \left({k\over N} \pi\right)\cos\alpha \right]+
{3}\pi^2\,{k\over N } {F\over\sqrt\lambda}\right). 
\label{fullaction}
}
The interpretation of the domain walls as infinitely
large spatial Wilson/'t Hooft
loops, makes it necessary to consider these 
boundary terms exactly as in 
\cite{ Hartnoll:2006hr, yamaguchi}.

The inclusion of this term is also essential for guaranteeing 
invariance under $k\to N-k$. The equation of motion \eqref{f}
implies that $F= - r^2 \sqrt\lambda \cos\alpha/2\pi$. Thus the
complete Lagrangian density in \eqref{fullaction} only depends on
$r^2$, and the action is minimized when $r=\pi T$.
The resulting formula for the tension is the same as the action for
5-branes computing Wilson loops in the antisymmetric tensor
representation,
\EQ
{
T_{F1} = {N\sqrt\lambda} {T^2\over 3} \sin^3\alpha.
\label{kstring}
}

\subsubsection{Tension of the 5-brane $k$-wall}

Above, we deduced the tension of the wrapped D5-brane carrying $k$
units of fundamental string charge. This object can be interpreted
as a domain wall associated to the breaking of a magnetic ${\mathbb
  Z}_N$ symmetry of hot $\N=4$ theory. S-duality on this yields 
the domain wall in the electric picture as a wrapped NS5-brane
carrying $k$ units of D-string charge.

Now we can S-dualize \eqref{kstring} by sending $g_s\rightarrow
1/g_s$, to obtain the tension of the
$k$-domain wall at strong coupling, interpolating between 
two generic ${\mathbb Z}_N$ vacua in the high temperature $\N=4$
theory:
\EQ
{
{\cal T}_k = \tfrac{4}{3}\pi \,N^2 {T^2\over \sqrt{g^2_{\rm
      YM}N}}\,\sin^3\alpha\,,\qquad  \quad
\cos\alpha\sin\alpha -(\alpha-\pi) = \tfrac{k}{N}\pi.
\label{strongtension}
}
Obviously, this bears little resemblance to the weakly coupled theory 
\eqref{weaktension}. Nevertheless, there are a few significant remarks
to be made. First, the dependence on the 't Hooft coupling is what one
expects from a weakly curved string dual (SUGRA), and it is
surprisingly in agreement with the weakly coupled Yang-Mills
description \eqref{weaktension}. The second interesting feature of the
tension at strong coupling is that when $k\sim {\cal O}(N)$, it 
scales as $N^2$, which is the scaling expected from a
classical soliton in a large $N$ theory (such as an NS5-brane in the
IIB dual) -- this feature also appears to be manifest at weak coupling
from the Casimir scaling of the tension. Finally, for $k\ll N$, the
strong coupling $k$-wall tension has an expansion in fractional powers
of $(k/N)$,
\EQ
{
{\cal T}_k \simeq 2 \pi^2 k\,N \,{T^2\over \sqrt \lambda}\left(1- C 
\left({k\over N}\right)^{2/3}+\ldots\right).
}

We round off our discussion of the strongly interacting theory with
one additional aspect of the realization of domain walls as D-branes
in the Type IIB theory. The $k=1$ wall is a D-string worldsheet,
pointlike on the transverse $S^5$. When $k\sim {\cal O}(N)$, the
D-strings expand into an NS5-brane wrapping an $S^4\subset S^5$. In
both cases, the respective probe branes spontaneously break the
$SO(6)$ isometry of the $S^5$ to an $SO(5)$ subgroup. While this
feature is obvious from the dual string perspective, its
interpretation in the gauge theory is somewhat obscure. In
particular, the spontaneous breaking of the global symmetry  
suggests that at least classically there are massless fluctuations of
the D-string worldsheet associated to fluctuations of a point, or an
$S^4$, in the $S^5$. These classically massless internal zero modes
are certainly not apparent in our weak-coupling ${\mathbb Z}_N$
instanton solutions, since the perturbative objects we discussed did
not have any scalar profiles turned on \footnote{In this context one
  should perhaps recall that if the effective domain 
wall theory were to be quantized, in the strongly coupled dual, the
apparently massless Goldstone boson fluctuations should be generically rendered
massive in the two dimensional world-volume theory of the domain wall.}. 

The direct connection between the domain walls and 't Hooft loops 
may help in further clarifying this issue. In particular, we know that
that the AdS/CFT correspondence does not provide us with direct access
to Wilson/'t Hooft loops of the $\N=4$ theory 
at strong coupling, but instead gives us the 
Wilson-Maldacena (or 't Hooft-Maldacena) loops
\cite{Maldacena:1998im,Rey:1998ik}
\EQ
{
W(C) = {1\over N}
\Tr \,{\cal P} e^{i\oint_C ds (A_\mu {\dot x}^\mu + \Phi_i \theta^i
  |\dot x|)}.
}
It would be interesting to compute the expectation values of these
kinds of loops at weak coupling and high temperature. In particular,
an interesting question is what kind of domain wall tension does the
't Hooft-Maldacena loop compute at high temperature, in the weakly
coupled theory and how it differs from the standard ${\mathbb
  Z}_N$ domain wall at weak coupling. Specifically, are there
classical solutions that have profiles for the scalars turned on, 
correlated in some way to the non-Abelian $A_0$ profile, providing the
solutions with internal zero modes.


\section{Conclusions}

In this paper we discussed the tension of domain walls in the
deconfined phase of ${\cal N}=4$ super Yang-Mills theory on ${\mathbb
  R}^3\times S^1$. While the
tension is proportional to $k(N-k)$ at the one and two loop level at
weak coupling, it exhibits a different behaviour at strong
coupling. This is expected since  already at three loops the
${\mathbb Z}_N$ domain wall tensions 
are not expected to exhibit a Casimir scaling. Nevertheless, a
quantitative comparison of 
the weak and strong coupling behaviours of the tensions reveals
intriguing features
as we see below.

In Figure \ref{graphs} we have plotted the Casimir scaling (weak coupling
behaviour) and the supergravity result (strong coupling) as a function
of $k/N$ for $N \rightarrow \infty$. The two graphs are normalized
such that ${\cal T}_{k/N=1/2}=1$. The maximum difference between the two
graphs is about $4\%$. 

\begin{figure}[ht]
\centerline{\includegraphics[width=4in]{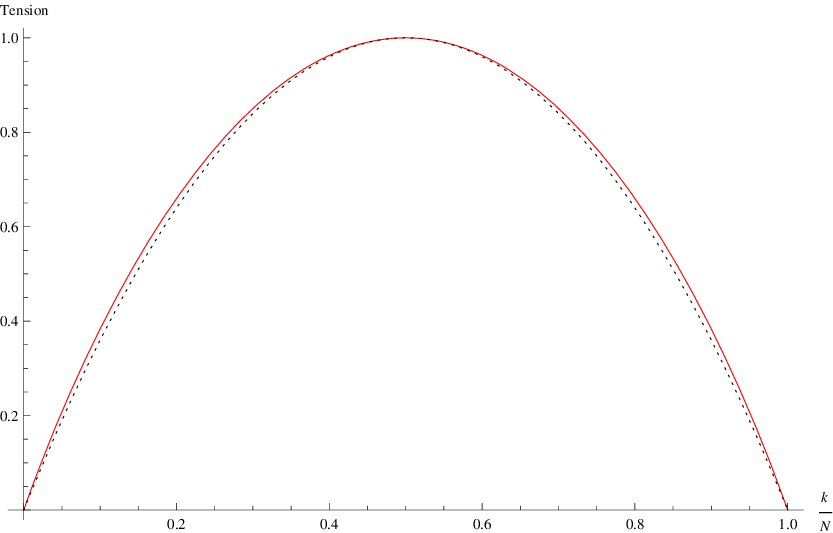}}
\caption{\footnotesize Casimir and $\sin ^3$ behaviours. The solid red
  line is $\sin ^3$ (strong coupling), whereas the dashed black line
  is the Casimir scaling (weak coupling).} \label{graphs} 
\end{figure}

Our results can be compared to lattice simulations, which were
performed for the pure YM theory. Within the measurement error, the lattice
results are compatible with a Casimir
scaling \cite{Bursa:2005yv,deForcrand:2005rg}, even at low temperature
(but above the deconfinement transition) where perturbation theory is
not applicable and there is no reason to expect an exact Casimir
scaling behaviour. It will be interesting to 
perform a more accurate simulation, in order to see a deviation from
Casimir scaling at low temperatures. Although the results of this
paper were obtained for ${\cal N}=4$ super Yang-Mills, they might shed
light on the expected tension of the pure Yang-Mills theory at strong
coupling (low-temperatures). Qualitatively, we expect from our study
that as the
temperature of pure Yang-Mills theory decreases the ratio of the
$k$-wall tension to the 
fundamental wall tension will increase, but by a very small amount
(see Figure \ref{graphs}).  

We conclude this paper with several open problems. It would be a
 useful
 exercise to
complete the two-loop calculation and to perform a three loops
calculation of the $k$-wall tension in ${\cal N}=4$ SYM.  Another open
question, relevant for both the pure and ${\cal N}=4$ SYM, is to
calculate the width of the domain wall in perturbation theory. We have
also mentioned that the domain walls at strong coupling appear, at
least classically, to have internal light degrees of freedom
naturally associated to a spontaneous breaking of the $SO(6)$ global
symmetry. Their appearance is closely related to the Wilson-Maldacena
loops in $\N=4$ theory, which are natural in the strong coupling
dual. It would be extremely interesting to firm up the connection between
these loops and ${\mathbb Z}_N$ 
domain wall solutions at weak coupling in hot $\N=4$ theory.  

Finally, it will be interesting to find the field theory that lives on
the ${\mathbb Z}_N$ domain walls. Since domain walls (at least at
large $N$) are believed to be QCD D-branes
\cite{Witten:1997ep}, there should be a 1+1 dimensional field theory
living on the domain walls, similarly to the Acharya-Vafa field theory
that lives on the domain walls of ${\cal N}=1$ SYM
\cite{Acharya:2001dz}. It would be fascinating to identify 
and show the existence of such a field theory, 
even in the case of pure YM theory. The answer to this question is 
certainly within reach for the $\N=4$ theory at strong coupling.

{\bf Acknowledgments:} We thank B. Lucini and C. Korthals-Altes for discussions. A.A. is supported by the STFC advanced fellowship award.

\startappendix
\Appendix{2-loop combinatorics}
\underline{\bf Case I: $C_a, C_b\neq 0$}: 
 $C_a$ and $C_b$ are only non-zero when $a$ and $b$ indices correspond
 to ladder generators. The remaining index on $f^{a,b,c}$ in 
 \eqref{eq:2loop_structure_expression} can either correspond to a
 Cartan (diagonal) generator or to a ladder operator. In the former
case
\begin{equation}
f^{ij,ji,{\rm diag}}f^{ij,ji,{\rm diag}}\,B_2(C_{ij})B_2(C_{ji})\sim 
2k(N-k) B_2(q)^2.
\end{equation}
 This result can be arrived at, by first noting that 
$i,j$ must be in separate sectors of $Y_k$ for $C_{ij}$ and $C_{ji}$
 not to vanish. It is then clear that there are $2 k(N-k)$ such terms
 which are identical since $B_2$ is an even function.

When the index $c$ corresponds to a ladder generator, carefully
following the combinatorics yields 
\begin{equation}
f^{il,lj,ji}f^{il,lj,ji}B_2(C_{il})B_2(C_{lj})\sim
\left[k(N-k)(N-k-1) + k(k-1)(N-k)\right]B_2(q)^2.
\end{equation}
\underline{\bf Case II: $C_a=0, C_b\neq 0$ and $ C_a\neq 0,C_b=0$ }:
Now either $a$ {\em or} $b$ can label a diagonal generator, 
or both $a$ {\em and} $b$ can correspond to 
off-diagonal generators. 
In the former case we have the contribution 
\SP
{
& f^{ij,\,{\rm diag},ji}f^{ij,\,{\rm diag},ji}
B_2(C_{ij})B_2(C_{\rm diag})\,\,\,\,\mathrm{or}\,\,\,\,
f^{{\rm diag},ij,ji}f^{{\rm diag},ij,ji}
B_2(C_{\rm diag})B_2(C_{ij})  \\\\
&\sim  4k(N-k)B_2(q)B_2(0).
}
Finally, when each of $a,b$ and $c$ map to off-diagonal, ladder generators
\begin{equation}
f^{il,lj,ji}f^{il,lj,ji}B_2(C_{il})B_2(C_{lj})
\end{equation}
 Here $i$ and $j$ are in different sectors, therefore forcing $l$ to
 be in a matching sector to one of them. Thus, either $C_{il}$ or
 $C_{lj}$ will vanish. Swapping sectors for $i$ and $j$ gives a factor
 of 2. 
\begin{equation}
\rightarrow 2\left[k(N-k)(N-k-1) + k(k-1)(N-k)\right]B_2(q)B_2(0)
\end{equation} 
\\
\underline{\bf Total of I \& II}:
 Summing all non-vanishing, $q$ dependent terms from both cases
 yields: 
\begin{eqnarray}
&\left(2k(N-k)+\left[k(N-k)(N-k-1) +
    k(k-1)(N-k)\right]\right) B_2(q)^2\nonumber\\ 
&+\left(4k(N-k)+2\left[k(N-k)(N-k-1) +
    k(k-1)(N-k)\right]\right)B_2(q)B_2(0)\nonumber\\ 
=&Nk(N-k)\left[B_2(q)^2+2B_2(q)B_2(0)\right]
\label{yangmills}
\end{eqnarray}
 Casimir scaling remains at 2-loops in pure Yang-Mills. 

There are two  main factors above which lead to the Casimir-like
 scaling: Firstly, the only 
 non-trivial $q$-dependence arises from either one, or both of $C_a$
 and $C_b$ being non-zero. Secondly, and more importantly, the combined
 function $B_2(C_a)B_2(C_b)$ is even. Thus for the cases where the
 indices of $f^2$, $a$,  $b$ and $c$ are off-diagonal (as in both of
 the cases above) the two contributions from $l$ being in different
 sectors sum to give the scaling. Explicitly, consider a general
 function of $C_a$ and $C_b$,  $H(C_a,C_b)$, where all $q$-dependence
 vanishes only for $H(0,0)$. Focusing on the analogous arguments to
 Case I: 
\begin{equation}
f^{ij,ji,{\rm diag}}f^{ij,ji,{\rm diag}}H(C_{ij},C_{ji}) = k(N-k)\left[H(q,-q) +
  H(-q,q)\right] 
\end{equation}
The diagonal contribution produces the Casimir scaling. However, for
the off-diagonal contributions: 
\begin{equation}
f^{il,lj,ji}f^{il,lj,ji}H(C_{il},C_{lj}) = k(N-k)\left[(N-k-1) H(q,-q)
  + (k-1) H(-q,q)\right]  
\end{equation}
If $H(C_a,C_b)$ is not an even function, there is a departure from
Casimir scaling. In particular, if $H(q,-q)= H(-q,q)$, then the two
loop result is proportional to $k(N-k)$.
The result, Eq.\eqref{yangmills}, is the pure Yang-Mills
result, however there is a plausibility argument that the Casimir
scaling remains at 2-loops for \nfour\, SYM.

\end{document}